# Enlightening the dark universe

Abhik Kumar Sanyal
**Department of Physics, Jangipur College, P.O. Jangipur, Dt.Murshidabad, Pin: 742213**

**Abstract:** Lot of avenues – the black hole, the wormhole, the dark matter, the dark energy etc., have been opened since the advent of General Theory of Relativity in 1915. Cosmology – the physics of creation and evolution of the universe, which was once thought to be beyond human perseverance, has now become a rich science of highest importance. However, the theory of gravitation – the oldest known interaction, is still not well understood. In the process of unveiling the evolutionary history of the universe, we shall explore some facts that suggest General Theory of Relativity is not the complete theory of gravitation.

1. **Introduction**:

The theory of gravitation was given by Sir Isaac Newton in 1690, with the simple understanding that gravity is mass. Law of gravitation being coupled to the equation of motion proved that there is nothing special in the planetary motion; rather Kepler's laws are true for any system obeying bounded central force motion. In fact it had even been attempted for atomic orbits by Rutherford, but failed since we now know that microscopic world is guided by a different physics altogether – the quantum mechanics. However Newton's law of gravitation remained unchallenged over centuries.

Out of the known four interactions, which are responsible for all physical phenomena observed in nature, Electromagnetism - presented in its final form by James Clark Maxwell in 1873, long before the advent of "Special Theory of Relativity" and "Quantum Mechanics" – "Quantum Field Theory" in particular, is very special. It is Lorentz invariant and so is compatible with Special Theory of Relativity. Further, it was suitably renormalized to obtain its quantum analogue - "Quantum electrodynamics". Further, unification with two other nuclear interactions – the weak and the strong, having quantum origin, had been found possible and they exist by the name of "Electro-Weak" interaction at $10^2$ GeV and "Grand unified theory" (GUT) at $10^9$ GeV. For some reason (discussed later) Einstein modified Newton's law of gravitation in 1915 and is commonly known by the name "General Theory of Relativity" (GTR), which essentially treats gravity as curvature of space-time. GTR leads to Newtonian gravity in the weak energy limit but fails in the very strong energy limit – close to the Planck's scale. Clearly, GTR should be replaced by a quantum theory of gravity near the Planck's scale. However, despite tremendous effort, all the attempts to construct a viable quantum theory of gravity went in vain. Thus unification with other interactions remained an illusion, and we have no knowledge of physics at or beyond the Planck's scale ($10^{19}$ GeV).

It is interesting to learn that gravity has been looked upon in different manner by Physicists working in different branches. To the General Relativists, the foundation of gravity is the Principle of equivalence guided by General Covariance, being described by a metric of space-time. To the Field theorist, Gravity is a massless Spin-2 field, plagued with ultraviolet (UV) catastrophe, guiding principle being the "unitarity". To the String theorist, stringy nature of gravitons makes scattering amplitudes very weak at Planck scale to control UV divergence. To all other Physicists, it is weak enough to neglect. Thus, gravity is fundamentally different from all other interactions, because it is the only interaction which is form invariant under general coordinate transformation, it is the only interaction described by spin - 2 field and it has uncontrolled UV divergence. In this sense, gravity – the oldest known interaction is still the most mysterious and not clearly understood. In this short review, we shall mainly study Cosmology to explore some of the fundamental problems associated with GTR.

2. **General Theory of Relativity-a brief overview:**

Newton's theory of gravitation remained unchallenged over centuries, because no experiment other than the perihelion precession of Mercury went against it. That the laws of physics remain invariant only in inertial frames bothered Einstein. Gravitation cannot be shielded and so inertial frames are obscure, except locally, since Galileo had shown that all the freely falling observers constitute local inertial frames. This is the law of equivalence – in disguise. However, the building block of a new theory of gravitation was the principle of "General Covariance" and the "Principle of Equivalence" [1]. The simplest tensor that is generally covariant was found by Einstein heuristically as $G_{\mu\nu} = R_{\mu\nu} - \frac{1}{2}g_{\mu\nu}R$, where, $R_{\mu\nu}$ is the Ricci tensor, $R$ is the Ricci scalar, $g_{\mu\nu}$ is the metric tensor, and $G_{\mu\nu}$ is called the Einstein's tensor. Since covariant derivative of the energy-momentum tensor vanishes in view of conservation law, so the equation cast by Einstein is

$$G_{\mu\nu} = R_{\mu\nu} - \tfrac{1}{2}g_{\mu\nu}R = 8\pi G\ T_{\mu\nu} \tag{1}$$

Above equation satisfies the Bianchi identity, $G^{\mu\nu}{}_{;\mu} = 0$ (semicolon stands for covariant derivative [1]) which is not an independent equation. The left hand side of equation (1) stands for curvature of space-time, while the right hand side denotes the type of matter content. Thus, gravity is treated as curvature instead of mass (figure 1). This equation immediately passed three very important tests, viz., perihelion precession of Mercury (43 arc sec that was due), bending of light (by Arthur Eddington) and Gravitational red-shift (by a very sensitive experiment performed at Harvard University). It also gives Newton's law of gravitation in the weak field limit. Due to symmetry in the indices, the above equation has ten components. Now Riemann tensor is given by

$$R^{\alpha}{}_{\beta\gamma\delta} = \Gamma^{\alpha}{}_{\beta\delta,\gamma} - \Gamma^{\alpha}{}_{\beta\gamma,\delta} + \Gamma^{\alpha}{}_{\mu\gamma}\Gamma^{\mu}{}_{\beta\delta} - \Gamma^{\alpha}{}_{\mu\delta}\Gamma^{\mu}{}_{\beta\gamma} \tag{2}$$

where, $\Gamma^{\alpha}{}_{\beta\gamma}$ are the Christoffel symbols and comma represents ordinary derivative. The metric is given as

$$ds^2 = g_{\mu\nu}dx^{\mu}dx^{\nu}. \tag{3}$$

Thus, given the metric tensor, which is used to find the Christoffel symbols using the relation,

$$\Gamma^{\mu}{}_{\beta\gamma} = \frac{1}{2}g^{\mu\alpha}\big(g_{\alpha\beta,\gamma} + g_{\alpha\gamma,\beta} - g_{\beta\gamma,\alpha}\big) \tag{4}$$

the Ricci tensor and the Ricci scalar are obtained under contraction of Riemann tensor and Ricci tensor respectively as $R_{\mu\nu} = R^{\alpha}{}_{\mu\alpha\nu}$, $R = g^{\mu\nu}R_{\mu\nu}$, and the left hand side of equation (1) is known. The energy momentum tensor is different for different fluid. For barotropic or non-barotropic ideal fluid, it is

$$T_{\mu\nu} = (\rho + p)u_{\mu}u_{\nu} + pg_{\mu\nu} \tag{5}$$

where, $\rho$ is the matter density and $p$ is the pressure, while, $u_{\mu}$ is the four velocity satisfying the relation $u_{\mu}u^{\mu} = -1$. The energy-momentum tensor for a scalar field is

$$T_{\mu\nu} = \phi_{,\mu}\phi_{,\nu} - \left(\frac{1}{2}\phi_{,\alpha}\phi^{,\alpha} + V(\phi)\right)g_{\mu\nu} \tag{6}$$

and so on. Hence, one knows the ten Einstein equations which are required to solve. If the metric coefficients are independent of time, then one deals with static space-time and probes astrophysical objects like stars, galaxies etc. On the contrary, if these are functions of time also, one probes the universe as a whole, which is studied in Cosmology. GTR opened new avenues like the black hole (figure 1), the wormhole (figure 2), the dark matter the dark energy etc. as solutions to the field equations.

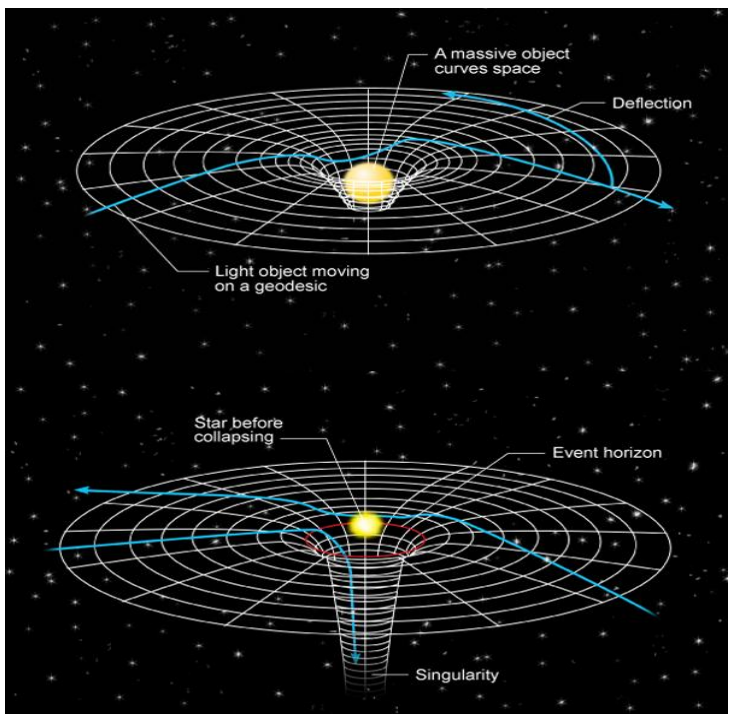

Figure 1: The figure on top depicts bending of light (blue) in the curvature, produced by a star. If the star is assumed to be the sun, other stars on the left, observed from the left hand side, are found at different locations in the sky when observed from the right hand side during solar eclipse. Taking images of hundreds of such stars, Eddington confirmed bending of light.

The figure (below) depicts gravitational collapse due to a dead star - the Schwarzschild's black hole (radius r, mass M). When light falls within the event horizon (r = 2M) – the region within the red circle - it can't escape. General Theory of Relativity administers a notorious unavoidable singular situation in the form of gravitational collapse at r = 0.

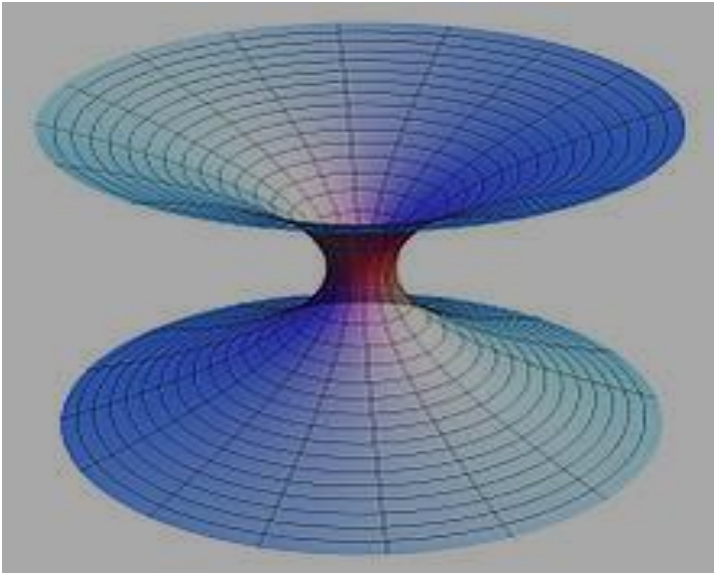

Figure 2: Two asymptotic flat or de-Sitter regions connected by a finite throat is called a Wormhole. Here, instead of the singularity, a throat appears avoiding gravitational collapse. Wormhole solutions are admissible in General theory of relativity, but it requires exotic matter for which the sum of the matter density and the pressure, $\rho + p < 0$.

### 3. Our Universe:

Astrophysics is the oldest science. In fact it has been developed since men could lie on their back. Nevertheless, common people still have no clear idea about the size of our universe. In fact in the early twentieth century, even specialist in this field had no clear idea in this respect. In this connection we may recall the famous debate between Harlow Shapley and Huber Curtis – the then two famous astrophysicists, on 16.4.1920. At that time, the Milky Way (figure 3) was thought to be the whole universe. However, many nebular clusters were detected by those times (figure 4 and 5) which are highly illuminated dust like objects.

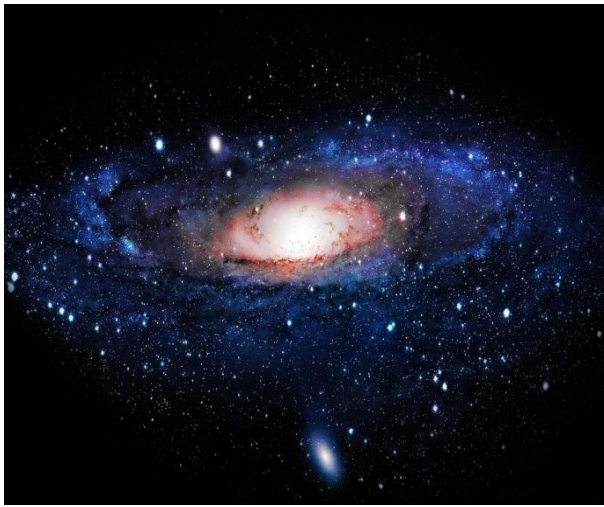
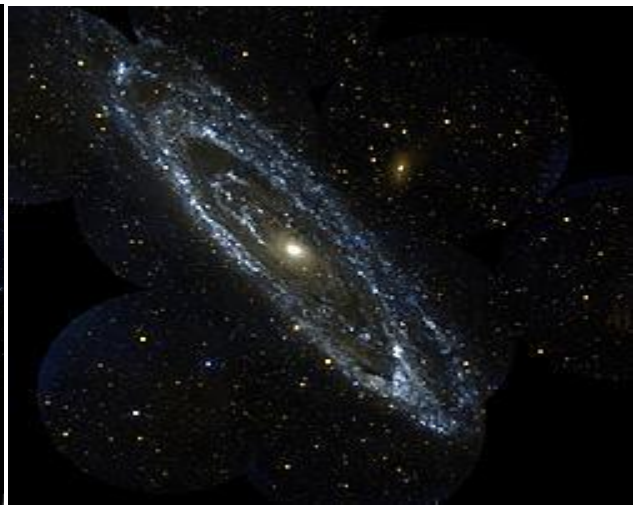

Figure 3: The Milky Way-thought to be the whole universe during early twentieth century.

Figure 4: The Andromeda nebula (nearest Galaxy).

In that debate, Harlow Shapley advocated his view that these nebular clusters are very much within our universe – the Milky Way. Huber Curtis on the contrary was advocating that these are far away from our universe and he named these as "Island Universes". In 1912, Henrietta Leavitt discovered a very important fact in connection with some very bright stars – the Cepheid Variables [2]. They send pulses with period between one day and fifty days depending on their absolute luminosities (table 1). Therefore these are treated as standard candles. Note that if one knows that there are 1000 Watt street lights, then it is possible to find the distances of various street lights by measuring their apparent luminosity ($L_a$) – the flux ($F$), using the following formula in Minkowski space-time

$$L_a = F = \frac{L_s}{4\pi d_L^2} \tag{7}$$

where, $d_L$ is called the luminosity distance. Between 1924 and 1925, Edwin Hubble measured the distance of different nebulae by observing hundreds of Cepheid Variable stars in different nebular clusters and concluded that these are far away from the Milky Way. He called these as galaxies and cluster of galaxies. So the universe extends much beyond the Milky Way. Presently, it is known that the universe contains 100 billion of galaxies and its size is about 10,000 Mpc, where Mpc stands for Mega parsec, while parsec is about 3.2 light years. Cosmology is the branch of physics which is related to the understanding of the creation and evolution of the universe, which we shall take up in the following section.

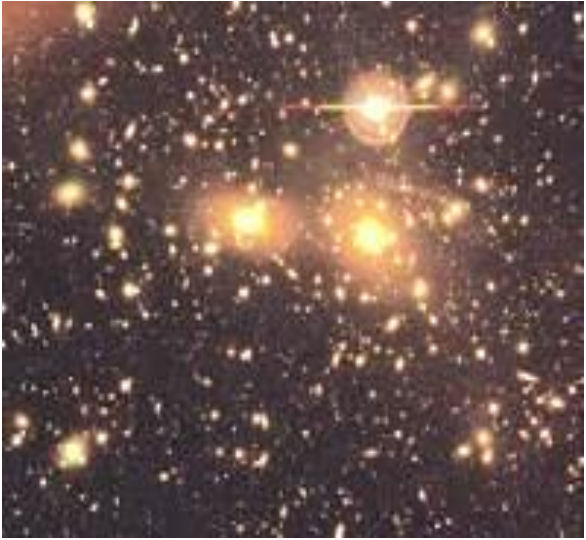

| Period ($T$) | Absolute Luminosity ($L_s$) |
|---|---|
| 1 day | $10^2$ Solar Luminosity |
| 50 days | $10^4$ Solar Luminosity |

Table 1: Period versus absolute luminosity of Cepheid Variable stars.

**Figure 5 The Coma cluster of nebulae**

### 4. Cosmology:

Although we observe stars, galaxies and cluster of galaxies all around, yet the universe is isotropic and homogeneous at scales greater than 100 Mpc – the reason (lies in the observed isotropy of CMB) will be explained later. Let us therefore start with cosmological principle that the universe is isotropic and homogeneous (inhomogeneity is treated as perturbation), which is described uniquely by the Robertson-Walker line element (c=1):

$$ds^2 = -dt^2 + a(t)^2 \left[\frac{dr^2}{1-kr^2} + r^2(d\theta^2 + \sin^2\theta \, d\phi^2)\right] \qquad (8)$$

Here, $a(t)$ – the scale factor is essentially the size of the universe, while the curvature parameter $k = 0$, $k = 1$, and $k = -1$ correspond to flat, spherical, and hyperbolic geometries respectively [1]. Note that the curvature of the universe is determined by the amount of energy and matter that is present in the universe. Einstein's equation therefore gives following two independent components

$$2\frac{\ddot{a}}{a} + \frac{\dot{a}^2}{a^2} + \frac{k}{a^2} = -8\pi G \, p; \qquad \frac{\dot{a}^2}{a^2} + \frac{k}{a^2} = \frac{8\pi G}{3}\rho \, . \qquad (9)$$

Additionally one has the Bianchi identity for the barotropic/non-barotropic fluid (5) under consideration, viz.

$$\dot{\rho} + 3H \, (\rho + p) = 0; \qquad H = \frac{\dot{a}}{a} \qquad (10)$$

which is not an independent equation, as already mentioned, rather may be found under suitable combination of Einstein's field equations (9). Nevertheless, it is very useful to solve the set of equations. Note that, we therefore have two equations to solve three field parameters, viz. $a(t)$, $\rho$ and $p$. Hence, one physically reasonable assumption is required. Aleksandre Aleksandrovich Friedmann (1922) for the first time assumed that the universe contains a gas in the form of pressure-less dust ($p = 0$) as we observe it today. Functional dependence of the scale factor with time indicates that the universe is expanding but decelerating, as $\ddot{a} < 0$. Universe must therefore had an early phase of evolution, when it was dominated by radiation ($p = \frac{1}{3}\rho$). The solutions he found although different but have the same feature. Jeorges Edourd Lemaitre in 1927 again confirmed the solutions independently. The solutions are

$$\text{Radiation:} \qquad p = \frac{1}{3}\rho; \qquad \rho a^4 = \rho_0; \qquad a = a_0 \, t^{\frac{1}{2}}. \qquad (11)$$

$$\text{Dust:} \qquad p = 0; \qquad \rho a^3 = \rho_0; \qquad a = a_0 t^{\frac{2}{3}}. \qquad (12)$$

Clearly, going back in time one finds that at, $t \to 0, a \to 0, \rho \to \infty, p \to \infty, R \to \infty$, indicating the expanding universe is plagued by an unavoidable initial singularity. We have already mentioned of the existence of Schwarzschild ($r = 0$) singularity appearing in static space-time [1]. Here again we encountered yet another singularity. Hence, one concludes that GTR is plagued with unavoidable singularity unless exotic matter dominates the very early universe.

### 4a. Attempt to remove the singularity by invoking darkness – the Cosmological constant.

Singularity is removed if it is pushed back to the infinite past. In the book "Confession" Archbishop St. Augustine (354A.D - 430A.D) asked the lord why did he create the universe (stating that "let there be light") on that particular day, why not earlier? Is there someone who urged him to do so? Then he would be the lord of lords. This was the first philosophical attempt to push back the creation to the infinite past. The problem of whether or not the Universe had a beginning was also a great concern to the German philosopher Immanuel Kant. He felt there were logical contradictions. If the universe had a beginning, why did it wait an infinite time before it began? He called it the thesis. On the other hand, if the universe had existed for ever, why did it take an infinite time to reach the present stage? He called it the antithesis. Both the thesis, and the antithesis, depended on Kant's assumption, along with almost everyone else, that time was absolute. That is, it went from the infinite past, to the infinite future, independently of any Universe that might or might not exist in this background. Einstein also did not support evolving universe scenario, to get rid of the creator god. If the universe is static – then creator god is removed – since it has never been created. So he modified his equation by introducing a term called the cosmological constant $\Lambda$, without giving any physical interpretation to it. The equation thus reads,

$$R_{\mu\nu} - \frac{1}{2} g_{\mu\nu} R + \Lambda g_{\mu\nu} = 8\pi G \, T_{\mu\nu} \tag{13}$$

A solution of the Field equations

$$2\frac{\ddot{a}}{a} + \frac{\dot{a}^2}{a^2} + \frac{k}{a^2} = -8\pi G \left(p - \frac{\Lambda}{8\pi G}\right), \quad \frac{\dot{a}^2}{a^2} + \frac{k}{a^2} = \frac{8\pi G}{3}\left(\rho + \frac{\Lambda}{8\pi G}\right) \tag{14}$$

for dust ($p = 0$) and $k = 1$ is then

$$\Lambda_E = 4\pi G\rho, \quad a_E = \frac{1}{\sqrt{4\pi G\rho}} = 10^{10} \text{ Light years.} \tag{15}$$

The radius of curvature of the universe is therefore a constant and the universe turns out to be static. But this solution was not accepted. We discuss the reasons below.

### 4b. The reason why such attempt failed.

i. Olber's paradox: J.P.L.Chesoaux (1744) and H.W.M Olber's (1826):Why the night sky is dark [3]?

If the universe is static and infinite in extent, then the number of stars between $r$ and $r + dr$ is $dn = 4\pi r^2 dr \, n$. Therefore, the radiant energy is $\int_0^\infty L_a dn \to \infty$. So, infinite static universe requires bright night sky. First possible answer to this paradox is that the interstellar medium absorbs the energy. But this reasoning may be ruled out immediately, since the interstellar medium is already in thermal equilibrium. The next possible answer is - since the stars are opaque, not all are observed. But every line of sight ends in a star, so this reasoning is also ruled out. The third option is that the universe is evolving (expanding) and is finite. Consider an observer in a boat sailing in the sea. Even if the sea is thickly populated with the boats, the observer will see only a finite number of those, if the boats are all moving away from him. Since most of them will be beyond the horizon. Movement of boats away from each other is equivalent to an expanding sea. Likewise, if the universe is expanding and is finite, light from a finite number of stars and galaxies reaches the observer and the paradox is resolved. So the solution to the Olber's paradox went against Einstein's static universe.

ii. The next reason that went against Einstein's static universe is the following.

Although the trace of equation (13) in vacuum ($p = \rho = 0$) is

$$R = 4\Lambda \tag{16}$$

meaning that the curvature scalar remains constant, nevertheless, the general solution gives exponentially expanding universe, viz.

$$R = 6\left(\frac{\ddot{a}}{a} + \frac{\dot{a}^2}{a^2} + \frac{k}{a^2}\right) \Rightarrow a = a_0 e^{\sqrt{\frac{\Lambda}{3}}\, t} \tag{17}$$

which is known as de-Sitter Solution, [1]. It clearly indicates an accelerating universe, since $\ddot{a} > 0$.

 iii. The final blow came from Edwin Hubble's (1929) discovery that the universe is expanding.

The light from distant stars and galaxies is not featureless, but has distinct spectral features characteristic of the atoms in the gases around the stars. When these spectra are examined, they are found to be shifted towards the red end of the spectrum. Balmer series of hydrogen spectrum has wavelengths 434.1, 486.1 and 656.3 nm for violet, blue and red lines respectively. For galaxies moving away with a velocity, $v = 0.1c$, where, c is the velocity of light in vacuum, the spectrum observed at wavelengths 479.8, 537.4 and 725.6 nm respectively. This shift is apparently a Doppler shift but is called cosmological red-shift since it indicates that all the galaxies are moving away from us, due to cosmic expansion. Using the results from the nearer ones, it becomes evident that the more distant galaxies are moving away from us faster. This is the kind of result one would expect for an expanding universe and this is what Hubble observed. The building up of methods for measuring distance to stars and galaxies led Hubble to find the fact that the red shift (recession speed) is proportional to distance, $v = H_0\, l$, $H_0 = 100h$ Km. s$^{-1}$Mpc$^{-1}$ = 9.78 $h^{-1}$Gyr, where $0.5 < h < 0.8$ is an uncertainty parameter involved in the measurement of $H_0$. If this proportionality (called Hubble's Law) holds true, it can be used as a distance measuring tool by itself. Let us first define proper physical distance between objects (galaxies) as

$$l(t) = a l_0 \rightarrow v = l\, H(t),\; H(t) = \frac{\dot{a}}{a} \tag{18}$$

If the universe is expanding, we can define a "Cosmological Redshift Parameter" $z$, which is the factor by which universe has expanded, since a photon is emitted from the source. If $\Delta\lambda$ is the observed change in wavelength corresponding to a known wavelength $\lambda$ received from a galaxy, then the expression for redshift is given by,

$$z = \frac{\Delta\lambda}{\lambda} = \left(1 + \frac{v}{c}\right)\left(1 - \frac{v}{c}\right)^{-1} - 1 \tag{19}$$

where, $v$ is the recessional velocity of galaxies. Hubble measured the change in wavelength of the Balmer lines of hydrogen received from Cepheid variable stars observed in galaxies. As an example, for the violet, $\lambda = 434.1\, nm$, and $\Delta\lambda = 45.7\, nm$. So, $z = 0.105275282$, $v = 0.0999761486c$. Hence, the recessional velocity of galaxies was measured, confirming the expansion of the universe. [Note that, Hubble's law is used to find the distance of the galaxies as, $l = \frac{v}{70\, Km.s^{-1}Mpc^{-1}} = 13 \times 10^8$ light years. In the process, redshift parameter measures the distance as well as the age of the universe, given a particular model]. In general, note that the luminosity distance must take into account the expansion of the universe and can be written in terms of the redshift, $z$ as

$$d_L^2 = a_0^2 r^2 (1+z)^2 \tag{20}$$

 iv. In 1930, Arthur Eddington informed Einstein that the static universe is unstable against small change in radius of curvature.

### 4c. The standard model

As George Gammow stated, Einstein discarded the idea of static universe and the cosmological constant term in 1931, stating it to be his greatest blunder in life. So the situation as it stood is listed below.

 a. Cosmological principle that the universe is isotropic and homogeneous has been treated as a fundamental principle. So the Robertson-Walker metric (8) corresponds to the fundamental symmetry of nature.
 b. Freidman model (solutions 11 and 12, corresponding to the field equations 9) is said to be the "Standard model of Cosmology" and is often referred to as FLRW model, by the name of Friedmann-Lameitre-Robertson-Walker [Note that if any other form of matter is incorporated in field equation 9, the model is no longer FLRW. This is a common error now-a-days].
 c. Initial singularity is called by the name "Big Bang", first given by David Todd Wilkinson.
 d. The present distance of an object which emitted light at the time of Big-Bang (this is approximate, since light was trapped for a while) is called the particle horizon,

$$r_{hor} = a_0 \int_0^{t_0} \frac{dt}{a} = 3\, t_0 = 2H_0^{-1}, \text{ i.e. } t_0 = \frac{2}{3}H_0^{-1} \tag{21}$$

where, $t_0$ is the age of the universe and $H_0$ is called the Hubble distance or the horizon. It is now known that the size of the observable universe is 10,000 Mpc containing 100 Billon galaxies and each galaxy contains 100 Billion stars. In the process, the darkness initiated by the cosmological constant term has been removed by the "Standard Model of Cosmology". The creator god has been reinstated with its divine grace and spiritual glory. However, the cosmological constant term $\Lambda$, which is related to the vacuum energy (the reason will be discussed in course of time), revived time and again with its darkness.

### 5. Success of standard model

What is Big-Bang? We shall try to give a reasonable answer afterwards. But at present, if we assume that at a very high temperature universe was created with a Big-Bang, and since then universe has expanded and cooled, then universe should be treated as the highest energy laboratory already available in nature. We can therefore present a brief summary of the evolutionary history of the universe, since after Planck's era ($10^{-43}$s) one can follow the evolution using the concepts of thermal physics and particle theory. Note that setting the Boltzmann constant $k_B=1$, the temperature is energy equivalent.

i. At $10^6$ GeV, hot Big-Bang begins.
ii. At 100 GeV, Electroweak phase transition occurs.
iii. At 100 MeV, Baryogenesis (Quarks-Hadron phase transition) occurs. Quarks condense under strong interaction to form Baryons (p & n).
iv. At 10 MeV, Thermal equilibrium amongst $\gamma$, $\nu$, e, ē, n, p is established.
v. At 0.1 MeV ($10^9$ K) < B.E of light nuclei, Nucleosynthesis took place.
vi. At 1eV matter energy density becomes equal to the radiation energy density.
vii. At 0.25eV atoms form and photons decouple forming CMB (universe became transparent and dark age started).
viii. At $10^{-3}$ eV first bound structure form and universe lightened up by the quasars (God perhaps started his job from here stating "let there be light").
ix. At $10^{-4}$ eV, which corresponds to 2.728 K, we are trying to explain nature.

**Nucleosynthesis and thereafter** [4]: When universe cooled down to $10^9$ K, i.e., less than the binding energy of light nuclei, neutrons and protons combine to form deuterons and deuterons combine to form helium nuclei following the reactions, n + p → $^2$D; $^2$D + $^2$D → $^4$He. As universe expands temperature falls and all $^2$D do not transform to $^4$He. Consequently, heavier isotopes could not form. Theoretical prediction of elemental abundance of light nuclei is 75% of $^1$H, 25% of $^4$He, .01% of $^2$D, $10^{-10}$ % of Li, has been confirmed experimentally. Further, baryon energy density (4%) calculated from elemental abundance of light nuclei and photon to baryon ratio (= $10^9$) have also been confirmed. At this stage, universe evolved like hot thick of soup of plasma containing e$^-$, e$^+$, p, n, $\gamma$, $\nu$. And as mentioned at around $10^2$s, after Big-Bang, (10 MeV), thermal equilibrium is established. Universe at that stage was opaque due to Thomson scattering between electrons and photons, following the reaction, $e + \gamma = e' + \gamma'; \lambda_1 \neq \lambda_2$, as depicted in figure 6.

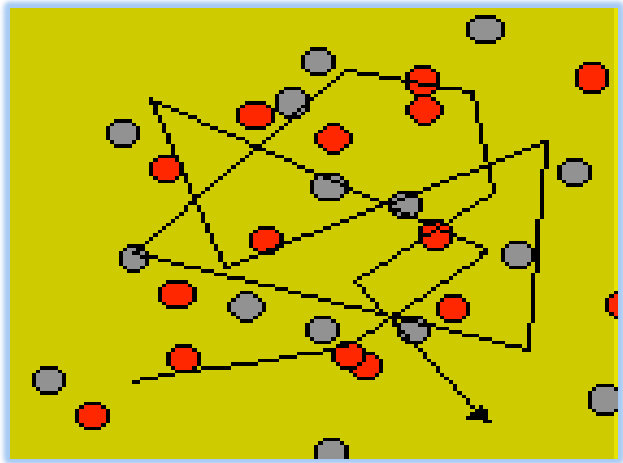

**Figure 6. The universe at this stage was opaque due to Thomson scattering between photons and electrons.**

Clearly the photons were trapped at this stage. At, T=3000K (0.25 eV) which is t = 4x10$^5$ years after the Big-Bang (according to standard model) photons decoupled and atoms started forming. This is known as the era of "Recombination". These photons since then, free stream all over, being cooled due to the expansion of the universe and must be found in the microwave region today. George Gammow first anticipated the existence of the black body spectrum of these photons at microwave region. Ralph Alpher and Robert Hermanin the year 1948, on the basis of Friedmann model predicted the temperature of these photons to be 5K. Gammow found the value to be 50K. This was due to wrong information regarding the present value of the Hubble parameter ($H_0$). Later, Robert Dicke and Yakov Zel'dovich (1960) followed by Dicke in 1964 found the temperature to be 3K. David Todd Wilkinson and Peter Roll were asked by Dicke to construct a so called Dicke Radiometer in the same year to detect this background radiation. In the meantime, Arno Penzias and Robert Woodrow Wilson made a Dicke Radiometer for Radio Astronomy and Satellite Communication. They found 7 cm background radiation having an excess of 3.5 K temperature in their Antenna in 1965. Having heard of their finding Dicke stated that "Boys we have been scooped". Penzias and Wilson won Nobel in 1978. This is now known as CMBR – The Cosmic Microwave Radiation Background [5], whose present value is 2.728±0.002K. This is the second (after Hubble's discovery of red-shift) astrophysical experimental result having cosmological consequence, since, the presence of CMBR confirmed unambiguously that the early universe was radiation dominated and the universe evolved to its late stage where the contribution of the radiation is insignificant and may be treated as containing pressure-less dust. In 1990 a satellite by the name COsmic Background Explorer (COBE) was launched to detect this background radiation. The Firas Radiometer detected the perfect blackbody spectrum as shown in figure 7 below [6].

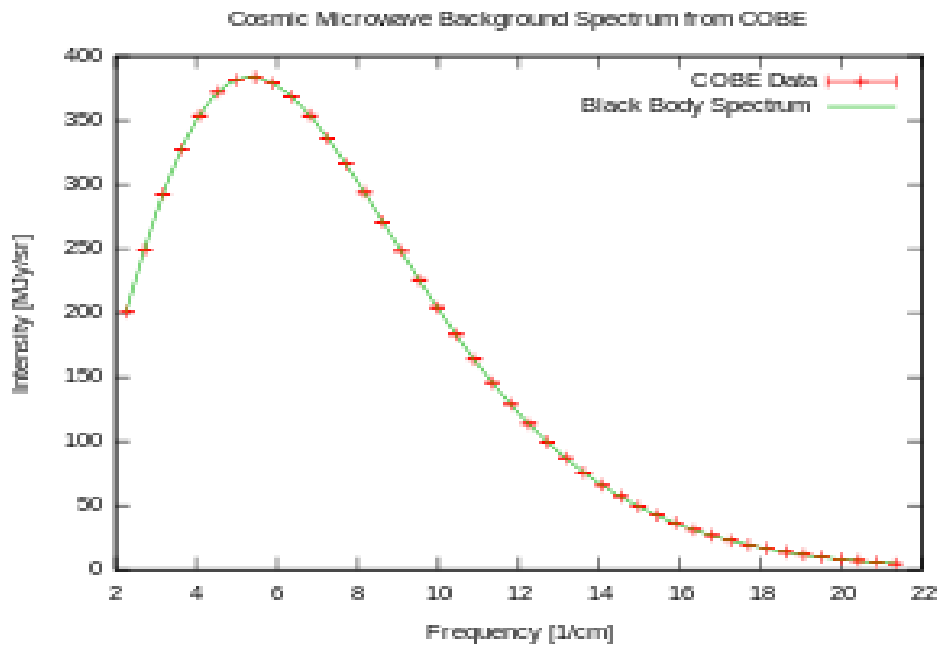

**Figure 7: Red points are experimental data obtained by Firas instrument set in COBE. This data curve unambiguously proved that the universe evolved from a Fire-Ball and Big-Bang has been confirmed.**

So, Big Bang cosmology, which was initiated in view of the FLRW model or more commonly the standard model, has been found successful in explaining cosmic evolution of the universe starting from a "Fire Ball of Plasma" containing neutrons, protons, electrons, positrons, photons etc. The theory of Nucleosynthesis followed by recombination (formation of atoms) has been accepted unambiguously after detecting Planck's black body spectrum. The detection of 2.7 K from all around the sky supports the cosmological principle that the universe is isotropic and homogeneous at large scale. However, at the small scale we observe in-homogeneities in the form of stars, galaxies and cluster of galaxies. In fact, COBE has also detected anisotropy in the CMB photons to one part in 10$^5$. These in-homogeneities can be explained by perturbing the Robertson-Walker metric. But standard model does not provide any answer to the seeds of perturbations. We shall come to the topic a little later, but before that let us discuss one more important findings of the last century.

### 6. The issue of Dark Matter:

Fritz Zwicky in 1933 observed large velocities of individual galaxies in COMA & VIRGO clusters. According to the Virial theorem, $K + \frac{U}{2} = 0$, where, $K = \frac{3}{2}M<v_r^2>$, $v_r$ being the velocity dispersion, and, $U = -\frac{GM^2}{R}$. Now taking

$R = 2Mpc$, Zwicky observed that the theorem holds provided, $M_{coma} \gg \sum M_{galaxies}$. In the above, $M$ stands for the mass. This means there exists substantial amount of non-luminous mass in the Coma cluster, apart from the total mass of luminous galaxies. This was not taken seriously at that time. But in 1970, peculiar velocities in the galaxies were observed by Peebles, Rubin and Freeman [7]. According to Kepler's law

$$\frac{GMm}{r^2} = \frac{mv^2}{r} \rightarrow v(r) = \sqrt{\frac{GM}{r}} \quad ie., \quad v(r) \propto \frac{1}{r}. \tag{22}$$

So, the velocity of stars in the individual galaxies must fall as the distance from the centre increases. However, the observation was different as shown in figure 8 below, which depicts that instead of falling off as Keplerian orbit the velocity remains almost constant at large distance from galactic centre, so that $M \propto r$. This confirms the presence of non-luminous matter – the dark matter in the galaxies, since otherwise the stars would have escaped.

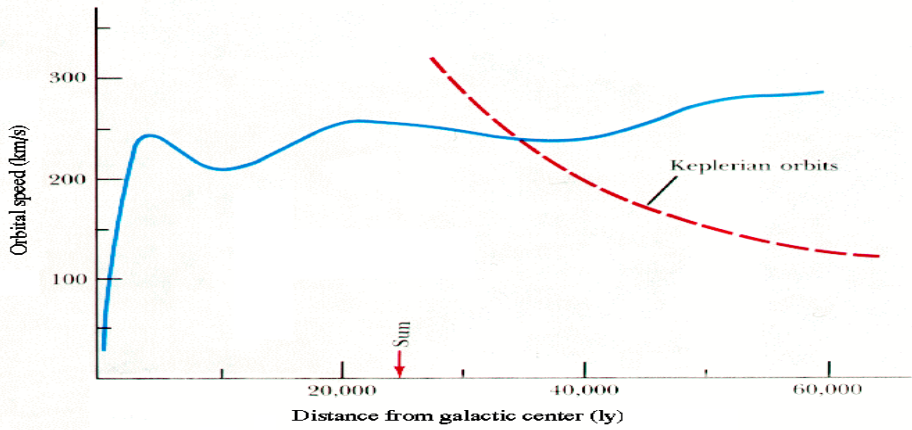

**Figure 8 Instead of following Keplerian orbit (red), the stars of individual galaxies move with almost constant velocity (blue), so that the mass is proportional to the distance from galactic centre.**

This has been tested over 1000 galaxies, observing 21 cm emission lines of neutral hydrogen instead of stars, which is much reliable. Now the question arises about the type of dark matter. For example there may be baryonic dark matter like the planets, the black holes etc. But the curve shows that the dark matter is evenly distributed. So the dark matter must be constituted of some non-baryonic particles which are non-interacting. One such known candidate is neutrino. These are relativistic particles and are called hot dark matter. But we will soon observe that structure formation (formation of galaxies and cluster of galaxies) also requires dark matter which must be non-relativistic (low speed), commonly known as cold dark matter (CDM). It has now been confirmed that radiation constitutes only $10^{-5}$%, while baryonic matter (luminous and non-luminous) constitute only 4% of the total energy density of the universe. So the next question "is dark matter constitute 96% of the total energy density of the universe"? We shall come to the answer in a while. But before that we shall try to understand the drawbacks of the standard model and its remedy.

7. **Problems with standard model:**

Already we know that GTR is plagued with singularity, since one can't avoid gravitational collapse without exotic matter. Apart from the singularity problem, the standard model is also associated with few more serious problems that we shall discuss in this section [8]. To better understand the problems we write the Friedmann equation as,

$$H^2 + \frac{k}{a^2} = \frac{8\pi G}{3} \rho \tag{23}$$

In the above, $H = \frac{\dot{a}}{a}$ measures the expansion rate of the universe. The critical density is the amount required to make the universe just flat and is found from the Friedmann equation, setting $k = 0$, as $\rho_c = \frac{3H^2}{8\pi G}$. The present value of the critical density is $\rho_{c0} = 2h^2 \times 10^{-29} g.cm^{-3}$. In relation to the critical density and also the density parameter $\Omega$, Friedmann equation (23) may also be expressed as

$$\frac{8\pi G}{3}(\rho - \rho_c) = \frac{k}{a^2} \xrightarrow{yields} (\Omega - 1) = \frac{k}{a^2 H^2}, \quad \Omega = \frac{\rho}{\rho_c}. \tag{24}$$

Clearly, for  $k = +1 \to \rho > \rho_c \to \Omega > 1$, the universe is said to be closed.
For  $k = -1 \to \rho < \rho_c \to \Omega < 1$, the universe is said to be open.
For  $k = 0 \to \rho = \rho_c \to \Omega = 1$, the universe is said to be flat.

### a. Flatness Problem:

In view of equation (24) one can find the following ratio (taking, $a \propto \sqrt{t}$, which is approximate, since at the late universe $a \propto t^{\frac{2}{3}}$, but it does not change the conclusion we draw) $\frac{(\Omega_{Pl}-1)}{(\Omega_0 - 1)} = \frac{t_{Pl}}{t_0} \approx 10^{-60}$, where, suffix "*Pl*" stands for Planck and we have taken $t_{Pl} = 10^{-43}$ sec, $t_0 = 13.7$ Gyr $\approx 10^{17}$ sec – the age of the universe. So,

$$\Omega_0 = 1 + (\Omega_{Pl} - 1) \times 10^{60} \tag{25}$$

It is apparent that slightest departure of $\Omega_{Pl}$ from 1, would have been magnified largely in the present value of $\Omega$ which is $\Omega_o$. If $\Omega_{pl}$ were a little over 1, universe would have been closed and get contracted long ago so that we could have lived. On the contrary, if it were a little less, then universe would have been expanded so fast, that structures could not have been formed. The universe where we live requires $\Omega_{Pl}$ close to 1 by $10^{-60}$ orders of magnitude. This is incredible, and is also called the problem of fine tuning. Who had made this very fine tuning of the density parameter at Planck's era, so that we might exist after 14 Billion years? Standard model does not find an answer to the problem.

### b. Horizon problem:

We have already noticed that CMBR is isotropic and homogeneous. The CMB photons received today from two different directions of the sky had never been in causal contact earlier, but still they show the same temperature. This is the horizon problem – how the CMB photons before coming within the horizon of each other, knew a priori that they should carry 2.7K temperature in the microwave region today? We can do some simple calculation to understand the situation more clearly. The temperature of the CMB photons falls with the expansion of the universe. As the universe expands adiabatically, the temperature cools, scaling as, $T \propto a^{-3(\gamma-1)}$. In the radiation dominated era $\gamma = \frac{4}{3}$, and so the temperature is inversely proportional to the scale factor. Hence we have

$$\frac{T_0}{T_D} = \frac{2.73}{3000} = \frac{a(t_D)}{a(t_0)} = \left(\frac{t_D}{t_0}\right)^{\frac{2}{3}}. \quad t_D = 2 \times 10^5 h^{-1} \, years \tag{26}$$

Where, suffix "0" as usual stands for today and suffix "*D*" stands for the era of decoupling. So the decoupling occurred nearly 4 x $10^5$ years after Big-Bang (taking $h = 0.7$). Therefore the distance travelled by CMB photons is

$$a(t_0) \int_{t_D}^{t_0} \frac{dt'}{a(t')} = 3t_0 \left[1 - \left(\frac{t_0}{t_D}\right)^{\frac{2}{3}}\right] \approx 3t_0 = 6000 h^{-1} Mpc \tag{27}$$

Last Scattering Surface is a sphere of radius, $d_H(t_0) = 6000 \, h^{-1}$Mpc. Now, the particle horizon at decoupling is $\approx 3t_D \approx 0.168 h^{-1} Mpc$, and today it is

$$0.168 h^{-1} \left(\frac{a(t_0)}{a(t_D)}\right) Mpc = 184 h^{-1} Mpc \tag{28}$$

Angle subtended by this decoupling horizon now is

$$\Theta_D = \frac{184}{6000} = 0.03^c \tag{29}$$

So the sky is split in $\frac{4\pi}{0.03^2} \approx 1.4 \times 10^4$ patches, which were never causally connected before emitting CMBR. Standard model does not have any answer to the question "why CMB photons received from $10^4$ causally disconnected patches of the sky show the same behavior"?

### c. Structure formation problem:

At present, on scales less than 10 Mpc, universe is unevenly distributed in the form of galaxies and cluster of galaxies. This must have occurred at early times due to gravitational instability, whence overdense regions became more overdense by attracting matter towards them. Given a perturbation $h_{\mu\nu}$ around the Minkowskian metric $\eta_{\mu\nu}$ as, $g_{\mu\nu} =$

$\eta_{\mu\nu} + h_{\mu\nu}$, one can perform some involved calculation to show how structures had formed. But standard model which tacitly assumed cosmological principle does not have any answer regarding the formation of the seeds of perturbation.

### d. Unwanted Relics:

Physics of early universe is described by particle physics which predicts creation of topological defects during phase transition as a consequence of symmetry breaking. If Big-Bang had started at T > $10^9$GeV, Gravitino (m=100GeV) would have survived, and Big-Bang-Nucleosynthesis (BBN) would not have been possible. Further, topological defects like monopoles (point like defects and would dominate the matter in the universe), cosmic strings (linear defects, characterized by some mass per unit length), domain walls (space divided into different regions having different phase being separated by walls of discontinuity described by a certain energy per unit area) were produced, which are not observed experimentally. Clearly, these are not the problems associated with the standard model, but particle physics expects that the standard model should answer why such defects are not observed in the present universe.

### e. Age problem:

Stars are known to be more than 13Gyr old, for example, HE 1523-0901 (A red giant in the Milky Way) is 13.2Gyr old. Now the inverse of present value of Hubble parameter $H_0^{-1}$ (the Hubble time) is related to the uncertainty parameter by the relation $H_0^{-1} = \frac{9.78}{h}$ Gyr, with $0.5 < h < 0.8$. In view of equation (21) the age of the universe turns out to be 13.04 Gyr taking the minimum value of $h$. However, most of the present observations [9] suggest that $h$ is close to 0.7, in which case, the universe is only 9.3Gyr old. So the stars are older than the universe. Standard model does not solve the issue.

## 8. Inflation-Darkness (Λ / φ) invited

All the problems discussed in previous section were resolved singlehandedly by invoking an Inflationary era in the very early universe close to the Planck's time, which was very short lived. Alan Guth [10] first suggested the inflationary scenario in 1984. Inflation in cosmology means very fast expansion of the universe, much faster than the speed of light. Note that it does not violate special theory of relativity, since no object as such moves with that speed. Inflation makes arbitrary curved universe – flat, solving the flatness problem. With inflation a causally connected region soon crosses the horizon. After inflation halts, the regions which went out enter the horizon. This means we can only observe a small part of the universe and the CMB photons which appear to come from different directions of the sky, were causally connected at very early epoch. This solves the horizon problem. Now, for such an inflationary scenario one requires an exponential ($a \propto e^{\Lambda t}$) or power law ($a \propto t^n, n > 1$) solution to the scale factor. Note that the field equations (9) may be cast as

$$\frac{\ddot{a}}{a} = -\frac{4\pi G}{3}(\rho + 3p) \qquad (30)$$

For inflation, we need accelerated expansion, $\ddot{a} > 0$, which is possible provided $\rho + 3p < 0$, so that the equation of state $w = \frac{p}{\rho} < -\frac{1}{3}$. This is not possible with baryonic matter and so one has to go beyond Friedmann solution, taking different type of energy source into account. For example, cosmological constant works, since $\rho_\Lambda + p_\Lambda = 0$, and so, $w = -1$. The solution as we already know is, $a = a_0 e^{\sqrt{\frac{\Lambda}{3}} t}$. Thus, had very early universe been vacuum dominated (baryonic matter being absent) in the presence of cosmological constant, accelerated expansion is realized and the universe inflates solving the flatness and horizon problem, while unwanted relics are inflated away. But, on one hand there is no seed of perturbation required for structure formation and on the other, such inflation never halts. As we already know that at some stage of cosmological evolution we require a fire ball – the hot thick plasma which gave birth to baryogenesis, nucleosynthesis and structure formation, so inflation must halt. This is called the graceful exit problem. So cosmological constant was abandoned once again and a scalar field was invoked by Alan Guth [10]. The scalar field equation is

$$\ddot{\phi} + 3H\dot{\phi} + V'(\phi) = 0 \qquad (31)$$

We understand that the effect of taking a cosmological constant term or a scalar field makes the vacuum non-trivial. The vacuum has weight. Guth assumed a scalar potential V(φ) having two minima. The scalar field being dubbed as inflaton is trapped in the false vacuum as shown in figure 9. The scalar field gets out of this local minimum by quantum tunnelling, after some characteristic time, which leads to bubble nucleation following first order phase transition. The bubbles experience a state of negative pressure. Once created, they continue to expand at an exponential rate and inflation is realized. However, this model of inflation also failed, since the bubbles expand very fast and the universe is left void of structure. In the process one again encounters the graceful exit problem. This is commonly known as the old inflationary model.

The problem was resolved by Linde and independently by Steinhardt and Albrecht in the *New Inflation* model [11] by assuming that the inflaton field evolves very slowly from its initial state, while undergoing a phase transition of second order. This is called slow roll condition requiring, $\dot{\phi} \approx 0$, so the equation of state, $w = \frac{p}{\rho} = \frac{\frac{1}{2}\dot{\phi}^2 - V}{\frac{1}{2}\dot{\phi}^2 + V} \approx -1$. In this model, one obtains ($N = \ln\frac{a(t_{end})}{a(t)} \approx M_{Pl}^{-2} \int_{\phi_{end}}^{\phi} \frac{V}{V'} d\phi \approx 65$ e-foldings) the amount of inflation just required to solve the flatness and horizon problems. After the universe cools to a critical temperature, the scalar field can proceed to its 'true' vacuum state energy without tunnelling. However, one of the largely accepted models is known as the

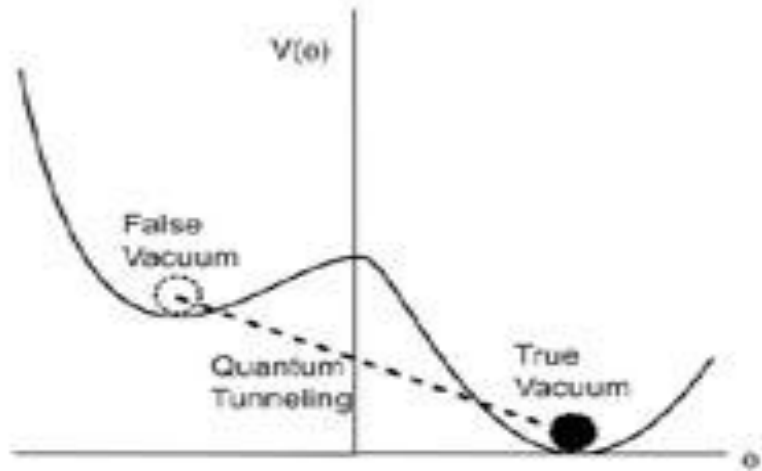

**Figure 9 The inflaton is trapped in a false minimum. It is freed from this minimum via quantum tunnelling resulting in a first order phase transition in the early universe. However, it suffers from graceful exit problem.**

*Chaotic Inflation* presented by Linde [12]. Here, no phase transitions occur. The inflaton is displaced from its true vacuum state by some arbitrary mechanism, perhaps quantum or thermal fluctuations, which then slowly rolls down the potential returning to the true vacuum. The universe thus undergoes inflation. Once the inflaton returns to its true vacuum state, the universe is reheated by the inflaton being coupled to other matter fields. After reheating, the evolution of the universe proceeds in agreement with the Standard Big Bang model. There are other models of inflation like the Extended, Hyperextended, Hybrid, Power law and the Intermediate inflationary models, which we shall not discuss here. However, Starobinski presented one important Inflationary model without phase transition [13], by modifying gravity under the addition of a curvature squared term ($R^2$) in the action. We just mention here, for the sake of completeness, that gauge invariant divergences make General Theory of Relativity non-renormalizable. So, in the very early universe, GTR must be replaced by an appropriate (unitary and renormalizable in 4-dimensions) quantum theory of gravity. So far, all such attempts require curvature squared term in the action, which automatically leads to an inflationary phase just after Planck's time. In this sense Starobinski's inflationary scenario is very important.

   a. **Seeds of Perturbation:**

We have so far tried to present a classical picture of inflation. Classically, inflation makes the scalar field homogeneous and isotropic on scales well inside the horizon. But inflation occurred in an epoch when all the fields were quantized. So quantum vacuum fluctuation of the field δφ(**x**,t) exists as the seeds of perturbation [8]. It leads to density perturbation δρ(**x**,t), as well as metric perturbation δg$_{\mu\nu}$, indicating the departure of homogeneity and isotropy. Additionally, there exists perturbation of each individual particle species δρ$_i$(**x**,t) and more complicated ones δΘ(**x**,t,**n**) specifying the number of photons at **x** with momentum along **n.** This is the CMB anisotropy. Finally, it also leads to the most important primordial curvature perturbation as, $R_k = -\left[\frac{H}{\dot{\phi}}\delta\phi_k\right]_t$. As inflation started, the perturbation went outside the horizon (since expansion rate of the universe is faster than the speed of light) and the curvature fluctuation freezes, since there is no causal effect. As inflation stops, light slowly captured the regions which went outside the horizon and in the process, the perturbations re-enters the horizon and causal connection is established. At this stage, there was a tussle between Gravitation and Random Particle motion. Gravitation tries to increase the density contrast δ$_k$= δρ/ ρ, attracting more matter in the over-dense region, but fails due to random particle motion, even in the presence of hot dark matter – the massless neutrinos. Thus, perturbation falls logarithmically and structure would not have been formed. Therefore, for the formation of structures, presence of non-relativistic Dark matter- the Cold Dark Matter (CDM) plays an important role. CDM wins over random particle motion growing the perturbation logarithmically so that a potential well is developed. Side by side, a tremendous pressure is created due to Thomson scattering between photons and electrons, so that the density contrast of the tightly coupled Baryon-Photon plasma fluid oscillates like a standing acoustic wave. Baryons fall in CDM potential well with same $\delta_k$, where,

$$\delta_k = \frac{\delta\rho}{\rho} = \frac{4}{9}\left(\frac{k}{aH}\right)^2 R_k, \text{ and, } \frac{\delta T}{T} = -\frac{1}{5}R_k|_{ls} \tag{32}$$

The above temperature contrast relation holds at large scale (few degrees) at the last scattering surface (a surface denoted by "*ls*" exists where photons suffered Thomson scattering for the last time before decoupling) is known as Sachs-Wolfe effect. As the temperature falls due to the expansion of the universe, electrons start forming atoms, Thomson scattering ceases and photons decouple to form CMB. Thus the pressure support is removed and acoustic oscillation stops, as photons start free streaming. The CMB anisotropy should therefore exhibit a series of acoustic or the so called Doppler peaks and troughs as a function of angular scale (small scale) providing a snap-shot of the acoustic oscillation.

Now, perturbations are of two types, viz., adiabatic and iso-curvature perturbations. If there is no change of entropy per particle for each species, so that fractional over-densities of each matter components (baryons, photons, and neutrinos) are the same, then it is said to be adiabatic perturbation. More precisely, in adiabatic perturbation, if in a spot over-densities of baryons and photons are 1%, then that of neutrino is also 1%. So the ratio is 1:2:3. On the contrary, in iso-curvature perturbation sum of the fractional over-density should vanish, since matter perturbations compensate radiation perturbations so that the total energy density remains unperturbed. That is, if baryon and photons have 1% over-density, neutrino must have 2% of under-density in the ratio 1:3:5. Now, Cosmic strings (the line like topological defects) was also found to explain all the problems of standard model. Effectively, it was treated as an alternative to the Inflationary scenario. Nevertheless, while Inflation predicts adiabatic perturbation, cosmic strings predict iso-curvature perturbation. We shall now briefly discuss the experimental results.

**b. Observations:**

With the advent of Inflationary scenario with its predictions, Cosmologists and Particle Physicists worked hand in hand for last three decades. As a result, different research organizations all over the world were enforced to enhance research budgets to an incredibly huge amount. The experiment started with the launch of COBE satellite in 1990. The prediction of CMB anisotropy was first observed by FIRAS instrument set in COBE satellite. It was then confirmed by Wilkinson Microwave Anisotropy Probe (WMAP), and recently by Planck (Figure 11). The angular resolution of Planck is sufficient to measure the distance between the hot (red) and cold (blue) regions at last scattering surface which confirms the present separation between galaxies. This clearly suggested that some seeds of perturbations (Vacuum fluctuation of scalar field) lead to the observed temperature anisotropy in CMB.

Further as mentioned, Inflationary scenario dictates that as it ends, the causally connected regions, which went outside the horizon, re-enters and must develop a standing acoustic wave in baryon-photon plasma fluid. Indeed a series of peaks and troughs – the Doppler peaks as functions of angular scale, have been found as a snap-shot of such acoustic oscillation. This is shown in Figure 11, which has been obtained analyzing data received from WMAP. The angle we see on the sky depends only on the angular diameter distance to the last scattering surface. The angular diameter distance depends on how light rays converge or diverge: in a spatially closed universe ($\Omega_k < 0$) they converge, and in a spatially open universe they diverge, relative to a flat universe. Primary information that comes from the angular location of the first acoustic peak is regarding the geometry of the universe. Therefore, in figure 11, the very first peak determines the curvature of space-time (angular scale). Second peak is much lower than the first. The maximum compression achieved by the masses at the bottom of the potential well depends on the pressure and the mass of baryons, and is larger for larger mass (because the masses have greater inertia). In contrast, the maximum rarefaction does not depend on the masses. So, if we fix everything but increase the baryon density, the compression peaks (first, third, fifth, etc.) would increase in height relative to the rarefaction peaks (second, fourth, sixths, etc.). As a result, the ratio of second to first peak amplitude tells us about $\Omega_b$, - the baryon density. In connection with other peaks, it is understood that in a radiation dominated universe, photons are redshifted, decaying the gravitational potential. Thus, temperature perturbation increases, enhancing the peaks. Since non-relativistic matter does not redshift, the phenomena is reversed. In reality both radiation and non-relativistic matter contributes. So relative ratio of peaks determine the total matter density parameter $\Omega_m$. Since, $\Omega_b$ is already known, so tracking the third (and more peaks in future experiments) one can have information about Dark matter. The ratio of the first, second and the third also signals Inflation with type of perturbation (angular scale).

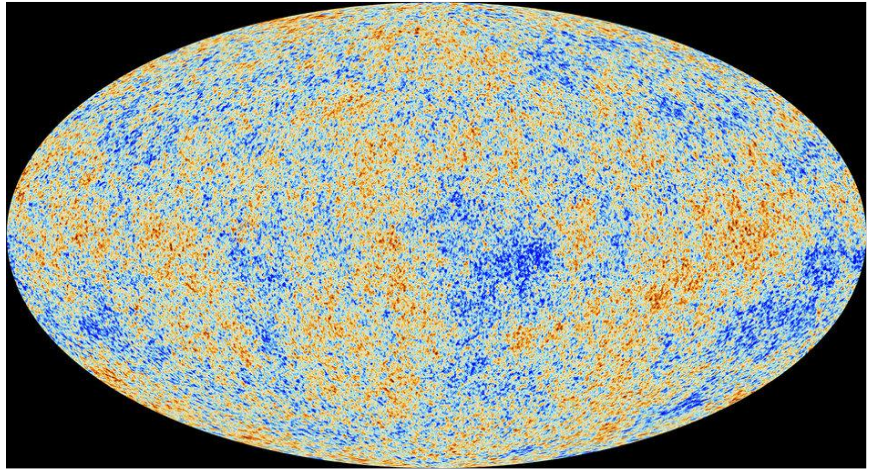

**Figure 10 Temperature fluctuation of CMB received from Planck at last scattering surface z = 1100.**

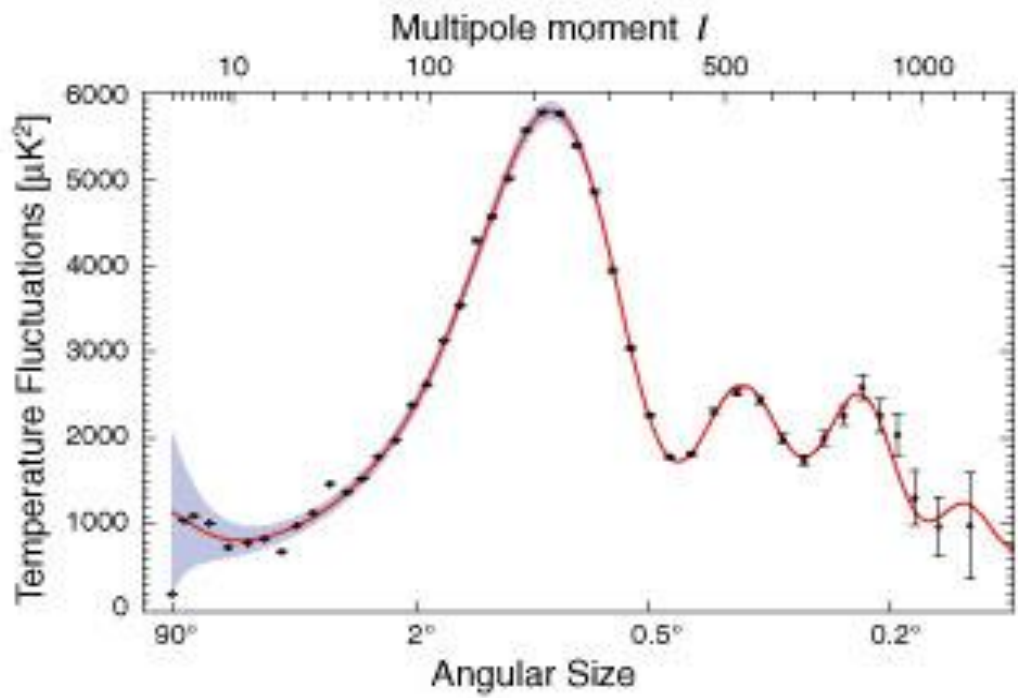

**Figure 11 The "angular spectrum" of the fluctuations in the WMAP full-sky map. This shows the relative brightness of the "spots" in the map vs. the size of the spots.**

### c. Experimental Data:

1. COBE detected anisotropy at $7^0$ angular scale, which is well outside the horizon at last scattering surface. Thus primordial form of perturbation detected.

2. BOOMERANG reported highest power fluctuation at less than $1^0$ angular scale. This gives the present value of the density parameter, $\Omega_0 = 1.02 \pm 0.04$, implying that the universe is nearly flat.

3. WMAP measured the peaks shown in fig. 11. From the first acoustic peak (at an angular scale of slightly less than $1^0$) WMAP tenth year data indicate that the density parameter corresponding to the curvature, $|\Omega_k| < 0.04$. This is remarkably close to zero, indicating that the universe is spatially flat to high precision. The ratio of the amplitudes of the second to the first peak tells us $\Omega_{b0} = 0.044$, which is consistent with the independent inference from big bang nucleosynthesis. Taking the third peak into account, the relative ratio of the peaks determine $\Omega m = 0.27$. Therefore, $\Omega_{DM} = 0.23$. This indicates that the density parameter amounting to 0.73 is still unknown.

4. DASI/VSA/CBI/QUAD/MAXIMA/ARCHEOPS/PLANCK with ACBAR precisely measured the ratio of first three peaks and found, $1^{st}:2^{nd}:3^{rd} = 1:2:3$, which confirms adiabatic perturbation.

5. SLOAN Digital Sky Survey (SDSS) mapped in detail one-quarter of the entire sky, determining the positions and absolute brightness of more than 100 million celestial objects. It also measured the distances to more than a million galaxies and quasars. In the process, it confirmed present value of the baryon density parameter to be $\Omega_{b0} \approx 0.4$.

The perturbations had close to Gaussian (i.e., maximally random) initial conditions [] and is adiabatic. So inflation has been confirmed, while cosmic strings are ruled out. Further, all the data suggested that universe contains only 4% of usual matter – the baryons, while the nature of the rest 96% is unknown and hence should be treated as darkness. Already we noticed that the seeds of perturbations can produce the presently observed structures by accreting matter in the over-dense regions in the presence of dark matter (weakly interacting) only. The present estimate is - about 23% of dark matter necessary for the purpose. Hence, rest 73% is still unknown.

So in a nutshell, the universe starts accelerating at around $10^{-42}$s i.e. just after Planck's time. Thus, once causally connected regions go beyond horizon. In the process, the seeds of perturbation arising from the scalar field also cross the horizon and perturbations freeze – meaning there is no interaction and perturbations are not disturbed. Inflation makes the universe flat (k → 0), while unwanted relics are inflated. Vacuum fluctuations of the scalar field φ create particles and the universe is reheated. Inflation stops at around $10^{-12}$ s and hot Big-Bang starts. Thus Big Bang is a hot thick soup of plasma and is not a singularity in any way. The standard model preceded by inflation pushed the singularity together with the Creator God to Planck's time. All the predictions of inflation and the standard model have been experimentally verified. The question that seemingly arises - is inflation a particular type of cosmological model, or an arbitrary constituent of any successful theory of the cosmos? Inflationary scenario involves quantized matter fields on a classical gravitational background. By assuming that the matter fields start out in a particular quantum state, the desired density fluctuation spectrum may be obtained. However, in the inflationary scenario, the question of initial conditions was largely ignored. So, to fully understand inflation, one needs a full quantum theory of gravity. Alas! A viable Quantum theory of Gravity, which can only answer what happened in the very early universe beyond Planck's time, is currently not available.

### 9. The Story of Late Universe

It was thought almost till the end of last century that other than the early universe, the cosmological evolution is explained thoroughly by the standard model. However, at least three issues were bothering cosmologists. The first one is the observation of ionized plasma (hydrogen and helium) in the intergalactic medium. After recombination there is no scope for re-ionization. There was a long dark age after free streaming of the photons. The universe was enlightened only after the quasars started forming. Most of the Astrophysicists are guided by the common idea that these quasars may possibly responsible for re-ionization [14], but the issue is debatable. Attempt to solve the problem by invoking higher order curvature invariant term ($R^{3/2}$), appearing in view of Noether symmetry, has also been made [15]. Nevertheless, this issue is far from being resolved. The second issue is regarding the age of the universe, which was left unsolved even after Inflationary scenario was proposed. The third issue is about the mismatch of the total matter density of the universe. We have already noted that the sum of baryonic and dark matter constitutes only about 30% of the total matter density of the universe. So around 70% was due. In fact it is possible that rest is also in the form of dark matter formed out of self-interacting gravitational field due to the little bit curvature left after inflation (since the curvature parameter $k \approx 0$, after inflation and may not be exactly equal). Theory of matter creation out of gravitational field had been explored thoroughly by Parker [16]. But, interesting result was revealed from COBE in the late 90's, analyzing Luminosity-Distance versus Redshift curve from SN1A supernova data. In the meantime WMAP was launched and two groups analyzed nearly 100's of SN1A data independently, before reaching the same conclusion that at present the universe is undergoing an accelerated stage of expansion [17] .WMAP worked till 2011 and data analysis from different groups also reached the same conclusion [18]. Like Cepheid variables, SN1A Supernovae are also treated as standard candles, but being much brighter, much larger distance may be probed (upto z ≈ 3). To understand the analysis, we take an example [19]. Let, *M* and *m* be the absolute magnitude (being related to log of $L_s$) and apparent magnitude (being related to log of *F* - Flux received). The luminosity – distance relation is then given as [1]

$$m - M = 5 \log_{10}\left(\frac{d_L}{Mpc}\right) + 25 \tag{33}$$

Now, let us take a pair of SN1A data: for *z* = 0.026, *m*= 16.08; while for *z* = 0.83, *m* = 24.32. So one can calculate the absolute magnitude of SN1A from the first data in view of equation (33) using the fact that $H_0 d_L = z$, for $z < 0.2$, and, $H_0^{-1} = \frac{3000}{h} Mpc$. The result is *M* = -19.09, and is the same (being standard candles) for all SN1A. Now using this value of absolute magnitude and the second data set of SN1A, one can find $d_L$ from equation (33) and hence $H_0 d_L$ = 1.16, where we have taken $h = 0.72$ [8]. One can also make theoretical estimate of $H_0 d_L$ using the following relation

$$d_L = \frac{L_s}{4\pi F} = \frac{1+z}{H_0} \int_0^z \frac{dz}{\sqrt{\Omega_{i0}(1+z)^{3(1+w_i)}}} \quad (34)$$

Where, the suffix $i$ denotes different components of fields. If one takes only matter (baryonic and dark) for which $w_i = w_m = 0$, then, $\Omega_m = 1$, and, $H_0 d_L = 0.94$, which does not match the experimental result. On the contrary, if dark energy in the form of cosmological constant is invited in addition, so that, $\Omega_m = 0.3$, $\Omega_\Lambda = 0.7$, while $w_m = 0$, $w_\Lambda = -1$, then the above integral yields $H_0 d_L = 1.19$, which agrees much better with the experimental result. This is just an example. Actual situation is illustrated in figure 12 below, which shows that the best theoretical fit with the experimental luminosity distance versus red-shift curve requires $\Omega_m = 0.3$, $\Omega_\Lambda = 0.7$ (green).

The puzzle with matter content in the universe is thus solved, since $\Omega_{total(0)} = \Omega_{m0} + \Omega_{CDM(0)} + \Omega_\Lambda = 1$. Note that, density parameter corresponding to radiation (luminous matter) contributes negligibly at the present epoch, since $\Omega_{R0} = 5 \times 10^{-5}$ (0.00005%) and inflation has made curvature density parameter $\Omega_k \approx 0$. This clearly implies that the energy-momentum tensor at the late stage of cosmic evolution should consist of baryonic matter, CDM, together with cosmological constant. As a result, one can calculate the so-called deceleration parameter corresponding to the field equations associated with Robertson-Walker metric containing different field components denoted by the suffix $i$, as

$$q = -\frac{a\ddot{a}}{\dot{a}^2} = -1 - \frac{\dot{H}}{H^2} = \frac{3}{2} \frac{\sum \Omega_{i0}(1+w_i)(1+z)^{3(1+w_i)}}{\sum \Omega_{i0}(1+z)^{3(1+w_i)}} \quad (35)$$

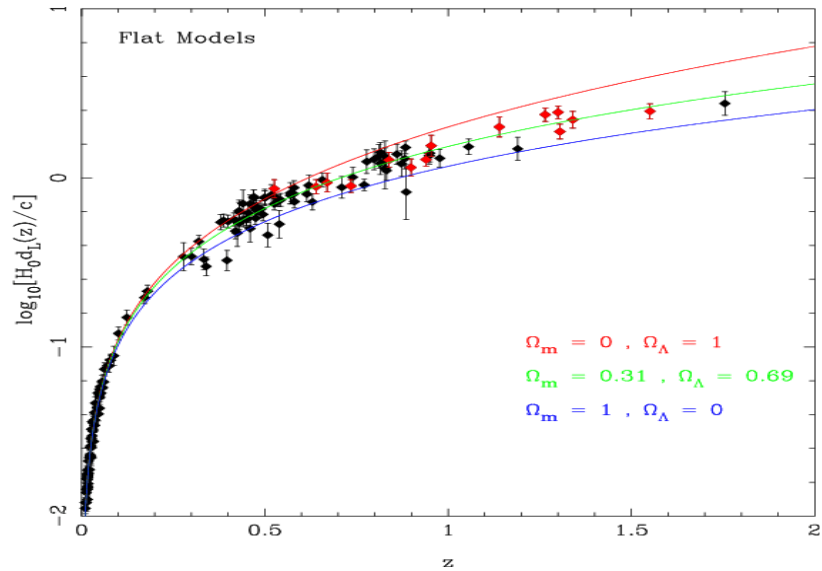

**Figure 12. SN1A data: Luminosity distance versus Red-shift data set fits with $\Omega_\Lambda$ =0.7, $\Omega_m$ =0.3 (courtesy Ref 18)**

The answer is, $q < 0 \rightarrow z = 0.7$, implying the universe has undergone an accelerated phase of expansion only recently as depicted in figure 13. So, acceleration is not what we observe, rather it is a bye product of the observed luminosity-distance versus redshift curve. Surprisingly an additional problem has been solved in the process – the so called age problem. Integral form of field equation in Robertson-Walker metric containing matter (baryonic and CDM) and cosmological constant, can be expressed further as

$$t_0 = \frac{1}{H_0} \int_0^\infty \frac{dz}{(1+z)\sqrt{\Omega_{mo}(1+z)^3 + \Omega_{\Lambda 0}}} \quad (36)$$

Taking, $\Omega_{\Lambda 0} = 0.7$ and $\Omega_{m0} = 0.3$ into account the age of the universe, $t_0 = 13.5$ Gyr. Thus the age problem is also solved invoking cosmological constant.

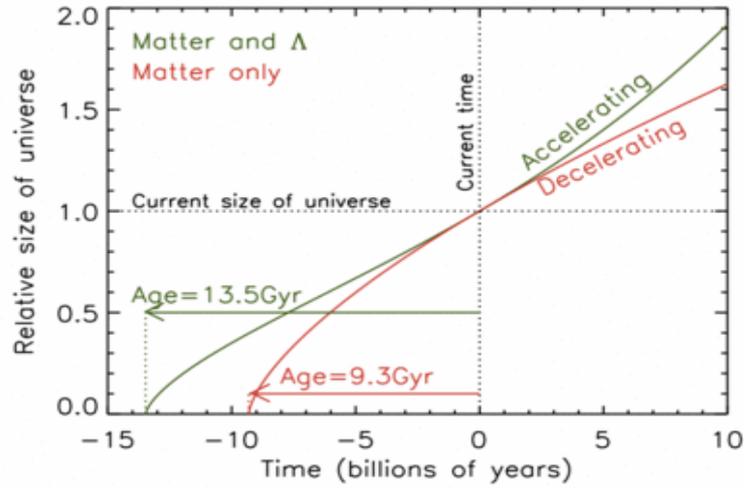

**Figure 13.** Red curve is for matter only, while green curve is plotted for 70% (Λ) and 30% matter, which shows recent acceleration and solves the age problem.

So, we conclude that the universe is mostly dark with the re-birth of cosmological constant Λ. The question therefore is "What is Λ – The Darkness?" Remember, Einstein inserted it by hand and thrown away stating it to his greatest blunder. It was invoked again in the context of Inflationary scenario and thrown away to resolve the issue of graceful exit. Once again cosmological constant reappears. But in the meantime, field theories in view of fundamental Physics, interpreted it as vacuum energy density ($\rho_{vac}$), i.e., sum of zero point energies of quantum fields with mass M. Simple calculation reveals

$$< T_{00} >_{vac} = \rho_{vac} = \frac{c^5}{G^2 \hbar} = 10^{74} GeV^4 \tag{37}$$

While, Friedmann equation reads: $3H^2 = 8\pi G \left(\rho + \frac{\Lambda}{8\pi G}\right)$. Clearly, Λ must be of the order of the present value of the Hubble's constant, i.e. $\Lambda \approx H_0^2$. So, $\rho_\Lambda = \frac{\Lambda}{8\pi G} \approx 10^{-47} GeV^4$. Therefore $\rho_\Lambda = 10^{-120} \rho_{vac}$. Hence, the energy-density corresponding to the cosmological constant necessary for solving the dark energy issue is 120 order of magnitude smaller than the vacuum energy-density. So, although cosmological constant, rather the so-called ΛCDM model gives the best fit to luminosity-distance versus redshift curve, solves the age problem and fits as well with all data obtained from WMAP, still it has been dropped once again. But remember, we now know that Λ has now received a physical meaning – it is the vacuum energy density - the weight of the vacuum. So, one has to answer why it is almost vanishing today. However, in order to resolve the puzzle of accelerating expansion of the universe and the dark energy issue, different models viz., Quintessence, K-essence, Hessence, Chaplygin Gas, Holographic dark energy, Brane induced models, Phantom fields, Tachyon fields, Scalar-tensor theory, Quintom model etc. were invoked, time to time [18]. In all these attempts, right hand side of Einstein's equation, i.e. energy-momentum tensor was modified and each model suffers from some sort of problems of their own. Additionally, solving the issue without invoking dark energy, by modifying the left hand side of Einstein's equation i.e. GTR has also been attempted. Examples are, Gauss-Bonnet (G) coupling, F(R) theory of gravity F(G) theory of gravity, i.e. functional dependence on the Ricci scalar or Gauss-Bonnet term etc. [20]. Particle creation phenomena by gravitational field are yet another option to resolve the issue taking 96% of dark matter into account [21]. In any case, there is practically no theory to explain or the cosmological constant, in disguise, is zero.

### 10. The very early universe:

Already we learnt that GTR cannot avoid gravitational collapse. So it is required to replace GTR by a viable quantum theory of gravity in the very early universe. The theory is viable in the sense that, it must be unitary, renormalizable and should produce GTR in the weak energy limit (note that success of GTR also lies in its Newtonian limit to pass solar test). Now, the very first criterion for quantization lies in the Hamiltonian formulation of a theory, which requires an action in the first place. Attempt to obtain the field equations of GTR from variational technique had serious problem. Under metric variation (variation with respect to $g_{\mu\nu}$) the following, so called, Einstein-Hilbert action

$$A = \int \sqrt{-g} d^4 x \left[\frac{R}{16\pi G} + L_m\right] \tag{38}$$

gives the field equations of GTR, provided both g_μν and its derivative are fixed at the boundary. This is illegal, since classical mechanics tells us that only the position or the field variable is kept fixed at the boundary and not their derivatives. In the above, R is the Ricci scalar and $L_m$ is the matter Lagrangian. This problem was resolved by York, followed by Gibbons and Hawking [22] by adding a surface term with the above action in the following form

$$A = \int \sqrt{-g} d^4x \left[\frac{R}{16\pi G} + L_m\right] + \frac{1}{8\pi G} \int_\Sigma \sqrt{h}\, K\, d^3x \tag{39}$$

which is the complete form of Einstein-Hilbert action. Here, $K$ is the trace of the extrinsic curvature tensor, being related to the derivative of the metric tensor. The boundary term gets cancelled with the one that appears under metric variation of action (38). The appearance of the surface term clearly depicts the difference of gravitational interaction from others. In the process, only six ($G_{11}, G_{22}, G_{33}, G_{12}, G_{13}$ and $G_{23}$) out of ten Einstein's field equations for GTR are retrieved, which are second order equations and so give dynamics. To retrieve other four ($G_{00}, G_{01}, G_{02}, G_{03}$) equations, it is required to find the Hamiltonian together with the three momenta, and set them to vanish. These four equation contain first derivative only and therefore do not give dynamics. Therefore, these are treated as constraints of the theory (the Hamiltonian constraint equation and the momentum constraint equations). On the contrary, one can make a 3+1 decomposition of the space-time metric (2) and express it as

$$ds^2 = -(N^2 - N_i N^i)dt^2 + 2N_i dx^i dt + g_{ij} dx^i dx^j \tag{40}$$

where, $N(t)$ is the lapse function and $N_i(\vec{x}, t)$ – the shift vector. The four constraints are now obtained under the variation of the lapse function and the shift vector and are therefore outcome of the so-called diffeomorphic invariance, being related to the general covariance – the building block of GTR. Thus Gravity may be looked upon as gauge theory. Canonical formulation of action (39) then follows and was presented by Arnwitt, Deser and Misner, being known by the name ADM formalism [23]. In fact, supplementing action (39) with a massive scalar field, Euclidean wormholes (representing quantum tunneling between different topologies) are administered, which avert singularity [24].

It has been mentioned that gauge invariant divergences make GTR non-renormalizabele. Stelle [25] had shown that a modified fourth order gravitational action,

$$A = \int \sqrt{g} d^4x \left[\frac{R}{16\pi G} + \beta R^2 + \gamma R_{\mu\nu} R^{\mu\nu}\right] \tag{41}$$

Although is renormalizable, contains five ghost degrees of freedom when expanded perturbatively around flat Minkowskian background, destroying unitarity. No attempt (String theory, M-theory, Supergravity) to formulate a complete quantum theory of gravity has so far been successful. So, in the absence of a complete quantum theory of gravity, quantization of action (39) was made in the cosmological context, to obtain certain insights, by the name quantum cosmology [26]. Since the Hamiltonian is constrained to vanish, so in the quantum domain the corresponding equation is obtained under consideration that the wave function is annihilated by the operator version,

$$\hat{H}|\Psi> = 0 \tag{42}$$

The above equation is called Wheeler-deWitt equation [25]. In the R-W metric above equation (42) may be expressed as

$$\left(\frac{\hbar^2}{12\, m_p^2\, a}\frac{\partial^2}{\partial a^2} - \frac{\hbar^2}{2a^3}\frac{\partial^2}{\partial \phi^2} - 3m_p^2\, ka + V(\phi)a^3\right)\Psi = 0 \tag{43}$$

Where, $m_p^2 = (8\pi G)^{-1}$ is the reduced Planck's mass and we have incorporated a scalar field. The absence of time now shows the difference from the Schrödinger equation. Therefore the standard probabilistic interpretation is unavailable. However, under suitable semi-classical approximation time reappears and it is understood that time is a semi-classical concept. The answer to the famous question "what was god doing before he created the universe?" is – "he was preparing a hell for those who will ask such question". Likewise, it is unfair to ask what happened before Planck's era – since; time loses its meaning beyond Planck's era within the periphery of GTR.

Since all the attempts to formulate a quantum theory of gravity contains curvature squared term in the action, so canonical formulation of the action (41) is necessary. Again the problem of boundary term appears. If only the scalar curvature squared term is taken into account, then apart from an additional surface term as shown below

$$A = \int \sqrt{-g} d^4x \left[\frac{R}{16\pi G} + \beta R^2\right] + \frac{1}{8\pi G} \int_\Sigma \sqrt{h}\, K\, d^3x + 4\beta \int_\Sigma \sqrt{h}\, K\, R\, d^3x \tag{44}$$

it requires fixing $\delta R = 0$ at the boundary [27]. This restricts classical solution by and large to de-Sitter or anti-de-Sitter solutions. Canonical quantization leads to a Schrödinger type equation, where an internal parameter (some power of the scale factor) acts as the time parameter [28]. Taking boundary term appropriately into account [29 - 32], the modified Wheeler-deWitt equation reads

$$i\hbar \frac{\partial \Psi}{\partial \alpha} = \widehat{H_e} |\Psi> \quad (45)$$

Where, $\alpha = a^3$ – the proper volume, acts as the time parameter and $\widehat{H_e}$ is the effective Hamiltonian operator which may be made self-adjoint under Weyl's operator ordering (complete symmetrization) and is therefore hermitian. Hence a unitary theory with standard probabilistic interpretation is administered. Further, under suitable semi-classical approximation the wave-function is found to be oscillatory being peaked around a classical inflationary solution [29 - 32]. Therefore, action (44) although is not renormalizable, appears to be more promising than Einstein-Hilbert action. Complete action (41) also shows the same bahaviour. One can now probe beyond Planck's scale and ask what happened when the scale factor is less than Planck's length $a < l_{Pl}$?

In the recent years, attempts have been made to explain late-time cosmic acceleration without dark energy by modifying theory of gravity. The corresponding action being

$$A = \int \sqrt{-g}\, d^4x\, F(R) + A_m = \int \sqrt{-g}\, d^4x \left(\frac{R}{16\pi G} + \beta R^2 + \gamma R^{-1}\right) + A_m + \Sigma \quad (46)$$

where, $A_m$ is the matter action and $\Sigma$ is the surface term. The beauty is - $R^2$ term dominates in the early universe giving inflation without phase transition, R dominates in the middle so that the standard model works and Big-Bang Neucleosynthesis, CMBR and structure formation are explained and $R^{-1}$ dominates at the late stage of cosmological evolution, so that late time acceleration is realized. Nevertheless, $R^{-1}$ has been set by hand and has no physical ground, since such a term is not produced by loop quantum gravity. On the other hand, $R^{3/2}$ term is realized from Noether symmetry of F(R) theory of gravity [33-36] and therefore, the following action is also well-suited for the same purpose, under appropriate choice of the coupling constant $\gamma$,

$$A = \int \sqrt{-g}\, d^4x\, [F(R) + L_m] = \int \sqrt{-g}\, d^4x \left(\frac{R}{16\pi G} + \beta R^2 + \gamma R^{\frac{3}{2}}\right) + \Sigma \quad (47)$$

Nevertheless, as already mentioned the additional condition required is to fix $R$ to be a constant at the boundary and so only de-Sitter or anti de-Sitter solution is realized. Therefore, the modified theory of gravity is useful to explain the early universe, but not at the late stage.

**Epilogue**: How much have we learnt and where do we stand then? We have addressed several issues in connection with the creation and the evolution of the universe. In the process, it has been made clear that GTR is applicable only in the weak energy limit. Therefore, at the very early universe, an appropriate (unitary and renormalizable) quantum theory of gravity should replace GTR. Since, modification of Einstein-Hilbert action by adding higher order curvature invariant terms have failed to resolve the issue, we can only hope that a modified version of superstring theory might resolve the problem in future. Until then, let the creator god be sited at the Planck's era. Another unresolved problem is the issue of cosmological constant - the vacuum energy density. Attempts to explain the vanishing of cosmological were made in view of wormhole physics [37] – but the problem is far from being resolved.

How far can we probe the universe? Not beyond the CMB (figure 10), since beyond CMB, light was trapped. However, all the data analysed from CMB so far, have confirmed an inflationary scenario just after Planck's epoch, followed by hot big-bang era, supporting the standard model of cosmology. Cosmological principle (that the universe is isotropic and homogeneous), which was initiated from a philosophical stand point has now been established scientifically. Thorough scientific explanation regarding structure formation (the stars, galaxies and the cluster of galaxies) is now available. Inflation is administered in view of GTR, provided a phi is available in the sky. Otherwise higher order correction to Einstein-Hilbert action is sufficient for the purpose, which becomes insignificant in the late era for giving way to the standard model.

Ionized Hydrogen and Helium are observed in the Inter-Galactic-Medium (IGM) suggesting that even after recombination (formation of atoms) there had been yet another phase of ionization – the "reionization". It is believed that earliest quasars are responsible for reionization. Nevertheless, it has been mentioned that higher order theory can also resolve the issue – since it acts as effective electromagnetic field [15]. This is still an unresolved issue.

Almost all attempts to fit luminosity distance versus redshift data curve end in accelerated expansion of the present universe. Nevertheless, for the purpose, one again requires scalar or Born-Infeld (tachyonic) type of fields, which act as dark energy. No experiment has yet been suggestive to detect dark energy. Modified theory of gravity can resolve the issue but there is a problem with the associated boundary term which restricts classical solutions to de-Sitter or anti de-Sitter. There is no clear explanation how a decelerating phase of expansion switched over to de-Sitter expansion without a phase transition in the middle. Further, such expansion is likely to end up in big-rip, when everything – even the atoms might tear apart. Philosophically, we are against such a destruction of the nature. Although cold dark matter (CDM) has not yet been detected, still we know that galactic rotation curves and structure formation require dark matter. Neutralino - a weakly interacting massive particle (WIMP) is a strong candidate in this regard. Therefore, a more acceptable possibility may be realized following particle creation phenomena out of gravitational field in the nearly flat universe [21]. Lot of experiments are being carried out in recent years to detect dark matter. If around 90% of dark matter is found, then the problem will be resolved without requiring dark energy or modifying gravity.

In any case, we come to the conclusion that GTR is not the complete theory of gravitation and a more rigorous theory is required to unveil nature.

Amidst the great universe/ the oceanic space and time/ oh lord! I wonder all alone.